\documentclass[aps,prl,twocolumn,tightenlines,superscriptaddress,
nofootinbib,preprintnumbers]{revtex4-1}

\usepackage{amsmath}
\usepackage{amssymb}
\usepackage{latexsym}
\usepackage{graphicx}
\usepackage{slashed}
\usepackage{pifont}
\usepackage{hyperref} 
\usepackage{color}
\usepackage{cleveref}
\usepackage{multirow}
\usepackage[usenames,dvipsnames]{xcolor}
\usepackage{soul}
\usepackage{cancel}
\usepackage{physics}
\usepackage{isotope}

\hypersetup{colorlinks=true,citecolor=[rgb]{0,0,0.8},
linkcolor=[rgb]{0,0,0.8},urlcolor=[rgb]{0,0,0.8}}

\definecolor{joelred}{rgb}{0.5,0,0}

\renewcommand{\vec}[1]{\mathbf{#1}}

\newcommand{\nxlo}[1]{%
   \ifnum0=#1\relax%
      \text{LO}%
   \else%
   \ifnum1=#1\relax%
      \text{NLO}%
   \else%
      \text{N}\ensuremath{^{#1}}\text{LO}%
   \fi
   \fi
}

\newcommand{\sdots}[2]{\vb*{\sigma}_{#1}\vdot\vb*{\sigma}_{#2}}

\begin{document}

\preprint{\hspace{13.45cm} MIT-CTP/4798} 

\title{Short Range Correlations and the EMC Effect in Effective Field
Theory}
\author{Jiunn-Wei Chen}
\email{jwc@phys.ntu.edu.tw}
\affiliation{Department of Physics, CTS and LeCosPA, National Taiwan
University, Taipei 10617, Taiwan}
\affiliation{Center for Theoretical Physics, Massachusetts Institute of
Technology, Cambridge, MA 02139, USA}
\author{William Detmold}
\email{wdetmold@mit.edu}
\affiliation{Center for Theoretical Physics, Massachusetts Institute of
Technology, Cambridge, MA 02139, USA}
\author{Joel E. Lynn}
\email{joel.lynn@gmail.com}
\affiliation{Institut f\"ur Kernphysik, 
Technische Universit\"at Darmstadt, 64289 Darmstadt, Germany}
\affiliation{ExtreMe Matter Institute EMMI, 
GSI Helmholtzzentrum f\"ur Schwerionenforschung GmbH, 64291 Darmstadt,
Germany}
\author{Achim Schwenk}
\email{schwenk@physik.tu-darmstadt.de}
\affiliation{Institut f\"ur Kernphysik, 
Technische Universit\"at Darmstadt, 64289 Darmstadt, Germany}
\affiliation{ExtreMe Matter Institute EMMI, 
GSI Helmholtzzentrum f\"ur Schwerionenforschung GmbH, 64291 Darmstadt,
Germany}
\affiliation{Max-Planck-Institut f\"ur Kernphysik, Saupfercheckweg 1, 
69117 Heidelberg, Germany}

\begin{abstract}
We show that the empirical linear relation between the magnitude of the
EMC effect in deep inelastic scattering on nuclei and the short range
correlation scaling factor $a_2$ extracted from high-energy
quasi-elastic scattering at $x\ge 1$ is a natural consequence of scale
separation and derive the relationship using effective field theory.
While the scaling factor $a_2$ is a ratio of nuclear matrix elements
that individually depend on the calculational scheme, we show that the ratio is
independent of this choice.  We perform Green's function Monte Carlo
calculations with both chiral and Argonne-Urbana potentials to verify
this and determine the scaling factors for light nuclei. The resulting
values for $^3$He and $^4$He are in good agreement with experimental
values. We also present results for $^9$Be and $^{12}$C extracted from
variational Monte Carlo calculations.
\end{abstract}

\maketitle

{\bf Introduction:} Deep Inelastic Scattering (DIS) of leptons on
hadrons can be precisely described as  high-energy (perturbative)
lepton-quark scattering weighted by the parton distribution functions
(PDFs) that describe the probability of finding a quark or gluon inside
the hadron. 
DIS has been used to map out the quark and gluon parton distributions
for the proton and subsequently nuclei.  In recent years, these
experiments have revealed new and intriguing glimpses of nuclear
structure that we seek to derive using effective field theory (EFT)
methods.

In 1983, the European Muon Collaboration \cite{Aubert:1983xm} measured
the structure functions $F^A_{2}(x,Q^2)$ describing DIS for iron and
deuterium targets, where Bjorken $x=Q^2/(2p\cdot q)$ and $Q^2=-q^2$ are
defined in terms of the target four-momentum $p$ and the momentum
transfer from the lepton to the target, $q$. 
The results of these experiments could not be explained by  nuclear
structure (i.e., momentum distribution of nucleons inside the nucleus)
without modifying the nucleon structure \cite{Aubert:1983xm}. 
This ``EMC effect" was unexpected since the typical binding energy per
nucleon is so much smaller ($<$1\%) than the nucleon mass and the energy
transfer involved in a DIS process.  The EMC effect has now  been mapped
out for DIS on targets ranging from  helium to lead (see
Refs.~\cite{Arneodo:1992wf,Geesaman:1995yd,Piller:1999wx,Norton:2003cb,
Hen:2013oha} for reviews)
and similar medium modifications of parton structure have been
investigated in other reactions \cite{Peng:1999tm,Norton:2003cb}.
The picture that has emerged is that the ratio
\begin{equation}\label{REMC}
R_\mathrm{EMC}(A,x)
=\frac{2F_{2}^{A}(x,Q^2)}{AF_{2}^{d}(x,Q^2)},
\end{equation}
with $A$ the atomic number and $d$ the deuteron, can
deviate from unity by up to 20\% over the range $0.05<x<0.7$. The ratio
has very little dependence on $Q^2$ and so we suppress it. Experimental
data also suggest that for an isoscalar nucleus, the $x$ and $A$
dependence of $R_{\rm EMC}-1$ is factorizable. That is, the shape of the
deviation of $R_\mathrm{EMC}$ from unity is independent of $A$ while the
magnitude of the deviation depends only on $A$
\cite{Date:1984ve,Frankfurt:1988nt,*Frankfurt:1981mk}. 
$R_\mathrm{EMC}$ forms a straight line in intermediate $x$, and one can
express the magnitude of the EMC effect by the  slope $d
R_\mathrm{EMC}(A,x)/d x$ for $0.35\le x\le 0.7$. 
Since Bjorken $x$ is  defined with respect to the parent nucleon of the
struck parton, it is bounded in the range $0\leq x \leq A$.

In recent experiments at Jefferson Lab, it was found that the ratio of
quasi-elastic (QE) scattering cross sections,
\begin{equation}\label{a2} 
a_2(A,x) \equiv
\left.\frac{2\sigma_{A}}{A\sigma_{d}}\right|_{1.5<x<2}, 
\end{equation}
forms an $x$-independent plateau with negligible $Q^2$ dependence for
targets from $^3$He to $^{197}$Au \cite{Fomin:2011ng,Hen:2012fm,Frankfurt:1993sp,Egiyan:2003vg,Egiyan:2005hs}.
This factor $a_2$ is referred to as the short range correlation (SRC)
scaling factor. 
A remarkable empirical discovery is that the EMC
slope and the SRC scaling factor $a_2$ are linearly related
\cite{Weinstein:2010rt,hen2012}. 

In this Letter, we explain this linear relationship using EFT and
compute $a_2$ in light nuclei. We first review the EFT description of
the EMC effect of Ref.~\cite{Chen:2004zx} which explained the
factorization of $x$ and $A$ dependence of $R_\mathrm{EMC}-1$, and then
show that the linear relation follows naturally from this.
Factorization also shows that, up to higher order corrections, $a_2$ is scheme and scale independent even
though it arises from scheme- and scale-dependent matrix elements in
different nuclei.  Finally, the values of  $a_2$ for $^3$He and $^4$He
are computed using the Green's function Monte Carlo (GFMC) method with
both chiral and Argonne-Urbana potentials to confirm the scheme and
scale independence and are compared with data, showing close agreement.
Results for $^9$Be and $^{12}$C extracted from variational
Monte Carlo (VMC) calculations~\cite{wiringa2014} are also discussed.

{\bf EFT Analysis:} Chiral EFT is constructed based on the chiral
symmetry of QCD. It has been successfully applied to many aspects
of meson \cite{Gasser:1983yg}, single \cite{Bernard:1995dp}, and
multi-nucleon systems
\cite{beane2012,*beane2001,*bedaque2002,*kubodera2004,*epelbaum2009,
*hammer2013}.
In particular, chiral EFT has been applied to PDFs in the meson and
single-nucleon
\cite{AS,CJ1,*CJ2,DMNRT,*DMT1,*DMT2,Detmold:2005pt,Hagler:2007xi,
Gockeler:2003jfa} and multi-nucleon sectors
\cite{Chen:2004zx,Beane:2004xf} as well as to other light-cone dominated
observables~\cite{CJ3,*Belitsky:2002jp,Chen:2003fp,Chen:2006gg,
Ando:2006sk,Diehl:2006ya}. 

The structure functions describing lepton-nucleus DIS,
$F_{2}^{A}(x,Q^2)$, can be expressed in terms of nuclear PDFs
$q_{i}^{A}(x,Q)$ (for simplicity of presentation, we choose the DIS
scheme where the renormalization and factorization scale are set equal
to the hard scale of DIS, $\mu=\mu_f=Q$, although the results below do
not depend on the scheme) as $F_{2}^{A}(x,Q^2)=\sum_{i}Q_{i}^{2}x\
q_{i}^{A}(x,Q)$, where the sum is over quarks and anti-quarks of flavor
$i$ of charge $\pm Q_i$ in a nucleus $A$.
 In what follows, we focus on the isoscalar PDFs, $q^A=q_u^A+q_d^A$; in
the relevant experiments, nuclear PDFs are typically ``corrected'' for
isospin asymmetry of the targets.
The dominant (leading-twist) parton distributions are determined by
target matrix elements of bilocal light-cone operators. Applying the
operator product expansion,  the Mellin moments of the parton
distributions,
\begin{equation}
\langle x^n\rangle_A(Q) = \int_{-A}^{A} x^n q_A(x,Q) dx, 
\end{equation}
are determined by matrix elements of local operators,
\begin{equation}
\langle A;p|\mathcal{O}^{\mu _{0}\cdots \mu _{n}} |A;p\rangle =
\langle x^n\rangle_A(Q)\, p^{(\mu_0}  \ldots p^{\mu_n)}
\end{equation}
with
\begin{equation}
\mathcal{O}^{\mu_{0}\cdots \mu _{n}}=
\overline{q}\gamma ^{(\mu_{0}}iD^{\mu _{1}}\cdots iD^{\mu _{n})}q ,
\label{O}
\end{equation}
where $(...)$ indicates that
enclosed indices have been symmetrized and made traceless and $D^{\mu
}=(\overrightarrow{D}^{\mu }-\overleftarrow{D}^{\mu })/2$ is the
covariant derivative. 

In nuclear matrix elements of these operators, there are other relevant momentum scales below $Q$:  $\Lambda \sim 0.5$ GeV is the range of
validity of the EFT, and $P\sim m_{\pi}$ is a typical  momentum inside the nucleus ($m_\pi$ is the pion mass). These scales satisfy $Q \gg \Lambda \gg P$ and the ratio $\Lambda/Q$ is the small expansion parameter in the twist expansion while the ratio $\epsilon \sim P/\Lambda \sim 0.2-0.3$ is the small expansion parameter for the  chiral expansion.

In EFT, each  of the QCD operators is matched to hadronic operators at scale $\Lambda$
 \cite{Chen:2004zx}
\begin{eqnarray} \label{matching}
\mathcal{O}^{\mu _{0}\ldots \mu _{n}} &\to& 
:\langle x^{n}\rangle _{N}
M^n v^{(\mu_{0}}\cdots v^{\mu _{n})}N^{\dagger }N
\left[ 1+\alpha _{n} 
N^{\dagger }N\right] \,,  \nonumber \\
&+& \langle x^{n}\rangle _{\pi} \pi^{\alpha} i\partial^{(\mu _{0}}
\cdots i\partial^{\mu _{n})} \pi^{\alpha}+\ldots: ,  
\end{eqnarray}%
where the operators enclosed by $: \  :$ are normal ordered (with respect to the vacuum state), 
$N\,(\pi)$ is the nucleon (pion) field, $v$ is the nucleon four
velocity  and $\left\langle x^{n}\right\rangle_{N(\pi)}$ is the $n$th
moment of the isoscalar quark PDF in a free nucleon (pion).
The $\left\langle x^{n}\right\rangle_{N\,(\pi)}$ terms are one-body
operators acting on a single hadron only, while the $\alpha_n$ terms are
two-body operators. Here we have only kept the SU(4) (spin and isospin)
singlet two-body operator $\propto\left( N^{\dagger }N\right) ^{2}$ and
neglected the SU(4) non-singlet operator $\propto ( N^\dagger
\mbox{\boldmath$\sigma$}N ) ^{2}-( N^{\dagger }\mbox{\boldmath$\tau$}N )
^{2}$ which changes sign when interchanging the spin
($\mbox{\boldmath$\sigma$}$) and isospin ($\mbox{\boldmath$\tau$}$)
matrices \cite{Mehen:1999qs}. The latter operator has an additional
$O(1/N_{c}^{2})\sim 0.1$ suppression in its prefactor \cite{KS} with
$N_{c}$ the number of colors.
We also replace the nucleon velocity by the nucleus velocity and include
the correction $i\partial/M$ to higher orders.

The relative importance of the hadronic operators of
Eq.~(\ref{matching}) in a nuclear matrix element can be systematically
estimated from the power counting of the EFT, which assigns a power of
the small expansion parameter $\epsilon$ to each Feynman diagram.  In
Weinberg's power counting scheme \cite{Park:1995pn}, the nucleon
one-body operator is $\mathcal{O}(\epsilon^{-3})$, the nucleon
two-body operator is $\mathcal{O}(\epsilon^{0})$, while the pion
one-body operator connecting two nucleons is
$\mathcal{O}(\epsilon^{n-1})$.  Since $\langle x^{n}\rangle _{\pi}=0$
for even $n$ due to charge conjugation symmetry, the $n=1$ pion
operator enters at $\mathcal{O}(\epsilon^{0})$ but for higher $n$ the
contributions either vanish or are higher order compared with the
other operators in Eq.~(\ref{matching}).

The same order of importance for these operators is also found using
the alternate power countings of Refs.
\cite{KSW981,KSW982,Valderrama:2014vra}, but with a less suppressed
two-body effect compared with the one-body nucleon operator.  Other
higher dimensional operators are omitted here because they are higher
order in the power counting \cite{Chen:2004zx}.

Using nucleon number conservation, 
$\langle A|:N^{\dagger }N:|A\rangle =A$, the nuclear matrix element of
Eq.~(\ref{matching}) for $n\ne1$ is
\begin{equation} \label{xA}
\langle x^{n}\rangle _{A}(Q)
=\langle x^{n}\rangle _{N}(Q)
\Bigl[A+\alpha_{n}(\Lambda,Q)\langle A|:(N^{\dagger}N)^{2}:|A
\rangle_\Lambda\Bigr],
\end{equation}
where $\alpha _{n}$ is $A$ independent but $\Lambda$ dependent and is
completely determined by the two-nucleon system. 
After an inverse Mellin transform, the isoscalar PDFs
satisfy
\begin{eqnarray}
\label{qA}
q_{A}(x,Q)/A
\simeq
q_N(x,Q)+g_{2}(A,\Lambda )\tilde{q}_{2}(x,Q,\Lambda),
\end{eqnarray}
where 
\begin{align}\label{g2}
g_{2}(A,\Lambda)={1\over 2A}\bigl\langle A|:(N^{\dagger}N)^{2}:
|A\bigr\rangle_\Lambda, 
\end{align}%
and $\tilde q_2(x,Q,\Lambda)$ is an unknown function independent of $A$.\footnote{The exception of $n=1$ in Eq.~(\ref{xA}) results from the relevant contribution of the pionic operator in that case. This implies that factorization is violated only for $x=0$ \cite{suppl}.}
This result also holds at the level
of the structure function \cite{Chen:2004zx},
\begin{eqnarray}
\label{F2A}
F_{2}^{A}(x,Q^{2})/A
\simeq
F_{2}^{N}(x,Q^{2})+g_{2}(A,\Lambda )
f_{2}(x,Q^{2},\Lambda) .\  \  \
\end{eqnarray}
The second term on the right-hand side of Eq.~(\ref{F2A}) is the nuclear
modification of the nucleon structure function $F_{2}^{N}$. The shape of
distortion, i.e., the $x$ dependence of $f_2$, which is due to physics
above the scale $\Lambda$, is $A$ independent and hence universal among
nuclei. The magnitude of distortion, $g_2$, which is due to physics
below the scale $\Lambda$, depends only on $A$ and $\Lambda$. 

{\bf Linear EMC-SRC relation in EFT:} 
At smaller $Q^2$, we can generalize the analysis in the previous section to all higher twist terms in the operator product expansion. 
For a higher twist operator $\mathcal{O}^{\mu _{0}\ldots \mu _{n}}$, its indexes need not to be symmetric nor traceless, but the matching is still similar to Eq.(\ref{matching}). The only difference is that chiral symmetry dictates that the pion one-body operator has at least two derivatives in the chiral limit
even if the operator has no index. For example, twist-three operators $G_{\alpha\beta}^2$ and $m_q \overline{q}q$ are matched to $(\partial \pi)^2$ and $m_{\pi}^2 \pi^2$ operators. Therefore the same power counting result holds to all orders in the twist expansion and we have
\begin{eqnarray}\label{sigmaA}
\sigma_{A}/A
\simeq
\sigma_{N}+g_{2}(A,\Lambda)\sigma_{2}(\Lambda),
\end{eqnarray}
where the $E$ (initial electron energy), $x$ and $Q^2$ dependence
of $\sigma_i$ is suppressed. 

With $\sigma_{N}$ vanishing for $x > 1$, Eqs.~(\ref{a2}) and (\ref{sigmaA}) imply
\begin{equation}\label{sigmax}
a_{2}(A,x>1)
\simeq
\frac{g_{2}(A,\Lambda )}{g_{2}(2,\Lambda)},
\end{equation}
for both DIS and QE kinematics
yielding a plateau in $a_{2}$ as observed experimentally at
$1.5<x<2$. (Fermi motion, an ${\cal O}(\epsilon)$ effect in the EFT,
extends the contribution of the single nucleon PDF to $x$ slightly above
1 so the onset of the plateau is also pushed to larger $x$.) Since
$a_{2}(A,x)$ is a ratio of physical quantities, it is independent of the
EFT cutoff scale $\Lambda$. The EFT analysis also predicts that the scale dependence of $g_{2}(A,\Lambda )$ is independent of $A$
as suggested in \cite{Furnstahl:2013dsa}. 

From Eqs.~(\ref{REMC}) and (\ref{F2A}), direct computation shows the
that 
\begin{eqnarray} 
\frac{dR_\mathrm{EMC}(A,x)}{dx}
\simeq
C(x)\left[a_{2}(A)-1 \right]
\label{EMC-SRC}
\end{eqnarray}
has a linear relation with $a_2$,
with $C(x) = g_{2}(2) [f_{2}^{\prime}F_2^N-f_2
F_2^{N^{\prime}}]/[F_{2}^N+g_{2}(2) f_2]^2$ independent of $A$ and
$\Lambda$ 
(here,
$f^\prime=df/dx$). 

{\bf SRC scaling factor:}
Short-range correlations in light nuclei have been examined
theoretically from several points of view
~\cite{feldmeier2011,wiringa2014,rios2014,Atti:2015eda,
weiss2015,Vanhalst:2014cqa,Bogner:2012zm}.
However, the focus of previous studies was on the one- or
two-body distribution functions in coordinate or momentum space, which
are scale and scheme dependent~\cite{Furnstahl:2001xq,Furnstahl:2010wd}.

Here we discuss their observable ratio, the SRC scaling factor,
Eq.~(\ref{sigmax}).
We calculate $a_2$ using the GFMC method, which is one of the most
accurate methods for solving the many-body Schr\"odinger equation for
nuclei up to $A \le 12$~\cite{Carlson:2014vla}. The GFMC method projects
out the lowest-energy state of a given Hamiltonian $H$ from a trial wave
function $\ket{\Psi_T}$ via the many-body imaginary-time Green's
function
\begin{equation}
\lim_{\tau\to\infty}e^{-H\tau}\ket{\Psi_T}\to\ket{\Psi_0},
\end{equation}
with $\tau$ the imaginary time and $\ket{\Psi_0}$ the exact many-body
ground state. A limitation of diffusion Monte Carlo methods is that
they require local potentials in practice, while nuclear forces
derived from chiral EFT are usually nonlocal. Recently local
chiral EFT interactions have been derived up to next-to-next-leading
order (N$^2$LO) in Weinberg power
counting~\cite{Gezerlis:2013ipa,Lynn:2014zia,gezerlis2014,Tews:2015ufa,
Lynn:2015jua}.
This enables us to use the GFMC method with chiral EFT as well as
phenomenological interactions
to study the scale and scheme
independence of $a_2$.

\begin{figure}[t]
\includegraphics[width=1.0\columnwidth]{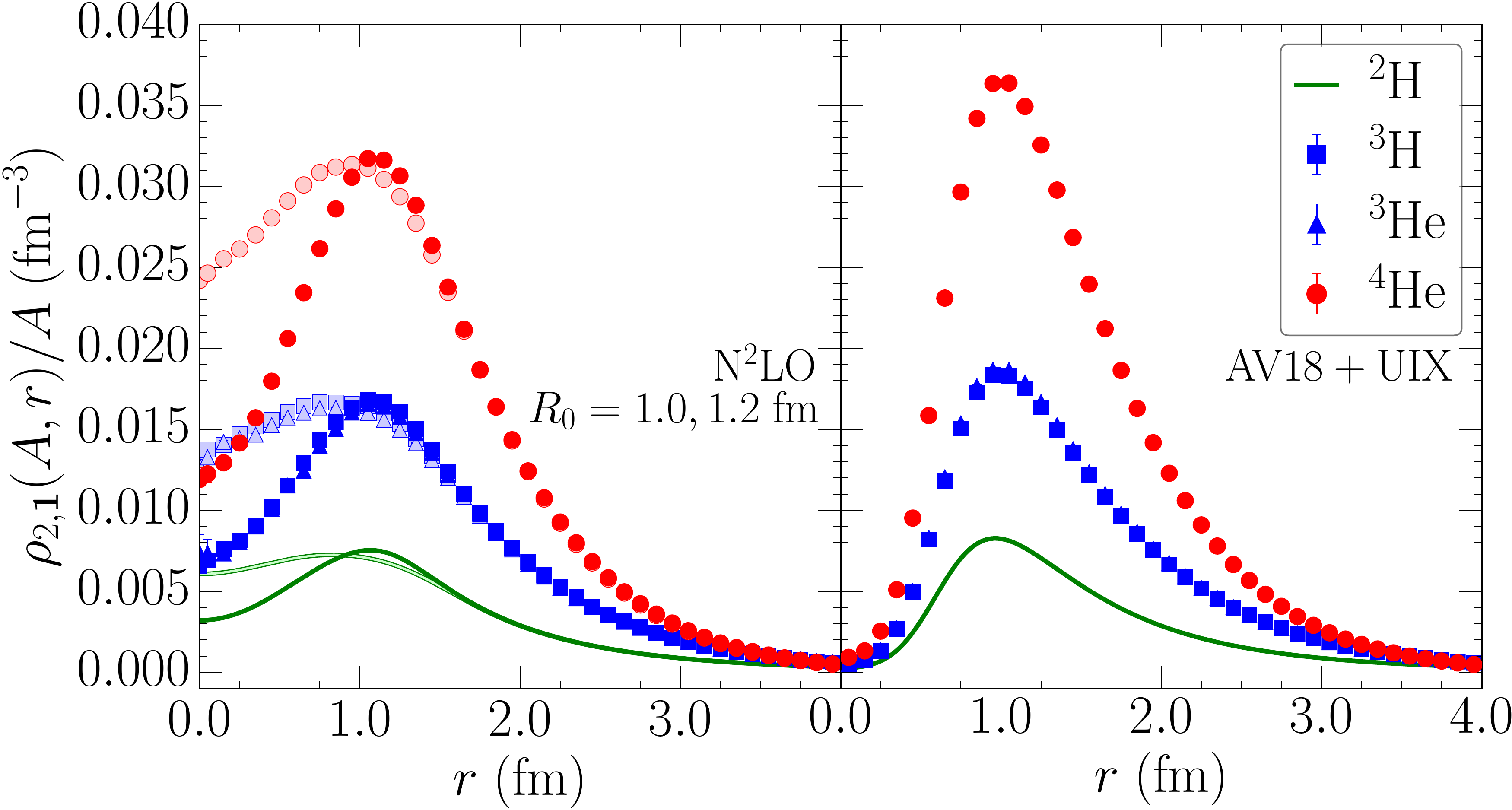}
\caption{\label{fig:rho2}
Scaled two-body distribution function $\rho_{2,\mathbf{1}}(A,r)/A$ for
$A=2,3,4$ nuclei as a function of relative separation $r$ for chiral 
interactions at N$^2$LO with two different
cutoffs (left panel) and for the AV18+UIX potentials (right panel). In
the left panel, the darker (lighter) points are for $R_0=1.0$~fm
($R_0=1.2$~fm). $A=2$ is solved exactly. For $A=3,4$ the error bars visible at small $r$ are
GFMC statistical uncertainties. The variation of the short-distance
behavior of the distributions shows clearly their scale and scheme
dependence.}
\end{figure}

\begin{figure*}[t]
\includegraphics[width=1.8\columnwidth]{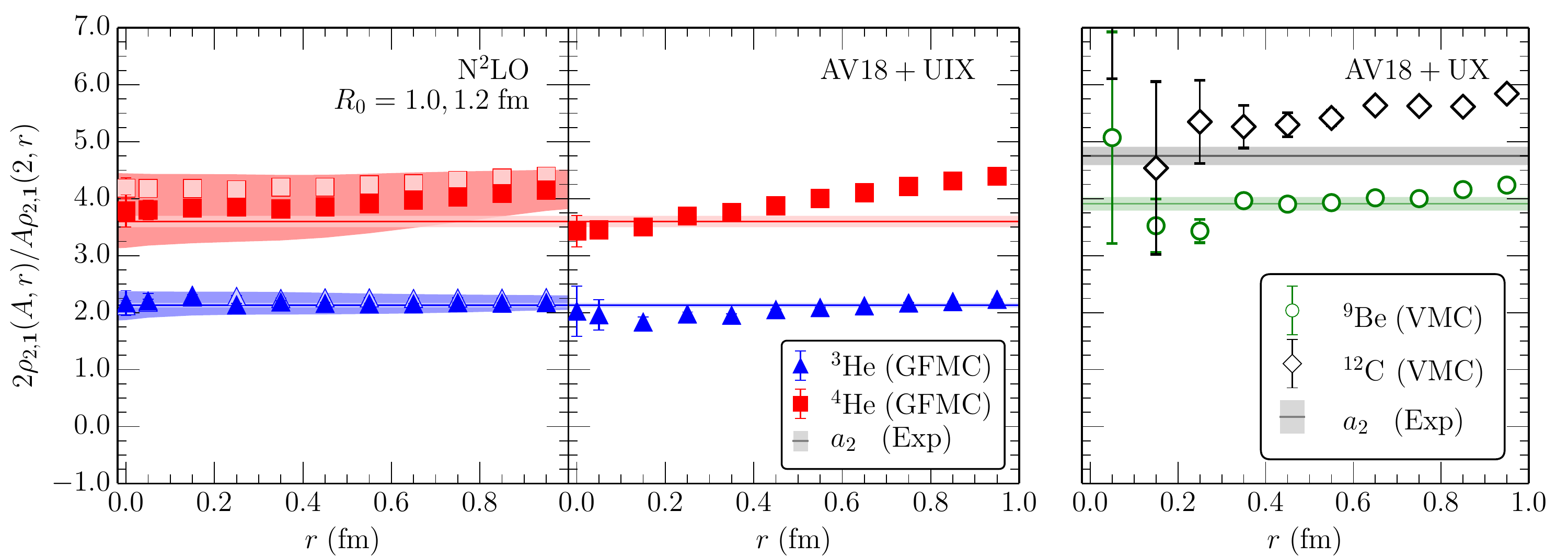}
\caption{\label{fig:a2}
Ratio of the two-body distribution functions for $^3$He (blue) and
$^4$He (red) to the two-body distribution function for the deuteron,
$2 \rho_{2,\mathbf{1}}(A,r)/ A \rho_{2,\mathbf{1}}(2,r)$, as a
function of relative separation $r$.  Results are shown for chiral 
interactions at N$^2$LO with two different
cutoffs (left panel) and for the AV18+UIX potentials (middle panel)
calculated using the GFMC method. In the left panel, the darker
(lighter) points are for $R_0=1.0$~fm ($R_0=1.2$~fm) and the bands
represent a combined uncertainty estimate from the truncation of the
chiral expansion added in quadrature to the GFMC statistical
uncertainties. The right panel shows the ratio for $^9$Be (green) and
$^{12}$C (black) for AV18+UX obtained from VMC
results~\cite{2bdens}. These ratios are compared to the experimental
values for $a_2$ from Ref.~\cite{hen2012}, given by the horizontal
lines.}
\end{figure*}

The function $g_2(A,\Lambda)$ of Eq.~(\ref{g2}) can
be obtained from the isoscalar two-body distribution
\begin{equation}
\label{eq:2bddist}
\rho_{2,\mathbf{1}}(A,r)=\frac{1}{4\pi r^2}
\Big\langle\Psi_0\Big|\sum_{i<j}^A\delta(r-|\vec{r}_i-\vec{r}_j|)
\Big|\Psi_0\Big\rangle,
\end{equation}
as a matrix element of a local operator, 
\begin{equation}\label{g2x}
g_{2}(A,\Lambda) = \rho _{2,\mathbf{1}}(A,r=0)/A.
\end{equation}
In Eq.~(\ref{eq:2bddist}), $\vec{r}_i$ is the position of the $i$th
nucleon and the sum runs over all pairs in the nucleus, so that the
integral over $\rho_{2,\mathbf{1}}(A,r)$ is normalized to $A(A-1)/2$.
We note that our EFT approach is not based on the experimentally
observed 
$np$-pair dominance~\cite{Subedi:2008zz,Hen:2014nza}, but
instead on the fact that, of the two $S$-wave two-nucleon operators, the
SU(4)-symmetric operator, $(N^\dagger N)^2$ (counting all pairs) is
dominant over the SU(4)-nonsymmetric operator (suppressed by a factor $\mathcal{O}(1/N_c^2) \sim 0.1$).
Thus, we include all pairs in Eq.~(\ref{eq:2bddist}).
Nevertheless, at short internucleon separations, we find a predominance
of $np$ pairs over $pp$ pairs by a factor $\sim 5$--10 for $^4$He and $^{12}$C.
In our GFMC calculations, the two-body distribution function is obtained
from a mixed estimate; for details see Ref.~\cite{Pudliner:1997ck}.

\Cref{fig:rho2} shows the scaled two-body distribution function
$\rho_{2,\mathbf{1}}(A,r)/A$ for $A=2,3,4$ nuclei for chiral two- and
three-nucleon interactions at N$^2$LO as well as for the
phenomenological Argonne $v_{18}$ (AV18)
two-nucleon~\cite{Wiringa:1994wb}
plus the UIX three-nucleon~\cite{Pieper:2001ap} potentials. 
The varying behavior of the two-body distributions at small separation
$r$ makes clear that $g_2(A,\Lambda)$ depends both on the scheme and
scale, where the latter is especially clear from the cutoff dependence
($R_0=1.0$~fm vs.~$R_0=1.2$~fm). Analogous to PDFs, one- and two-body
distribution functions depend on the renormalization scheme and
scale and hence are not physical quantities~\cite{Furnstahl:2010wd}.
However, the factorization derived
in EFT shows the ratio $a_2$ should be scheme and scale
independent.

Using Eqs. (\ref{sigmax}) and (\ref{g2x}), $a_2$ 
is obtained from the ratio
\begin{equation}
\label{eq:gfmca2}
a_2 
\simeq
 \lim_{r\to0}\frac{2\, \rho_{2,\mathbf{1}}(A,r)}
{A\, \rho_{2,\mathbf{1}}(2,r)},
\end{equation}
where we calculate the behavior at $r=0$ by linearly extrapolating from
the smallest two $r$ values to zero separation.
In EFT, locality only means a shorter distance than the
resolution scale. Hence, we expect one can replace $r\to0$ in
Eq.~(\ref{eq:gfmca2}) by smearing within $r<R$ 
(a scale set by, but not necessarily equal to, $R_0$), 
and still get the same $a_2$.

We see indeed this is the case in Fig.~\ref{fig:a2}. 
The left two panels show $a_2$ for
$^3$He and $^4$He calculated using the GFMC method with the chiral
N$^2$LO interactions and for the phenomenological AV18+UIX
potentials. The right panel shows results extracted from VMC
calculations~\cite{2bdens} for the AV18+UX potentials for $^9$Be and
$^{12}$C. The red and blue bands in the left panel represent a
combined uncertainty estimate from the truncation of the chiral
expansion~\cite{ekm2015} added in quadrature to the GFMC statistical
uncertainties. The ${\cal O}(\epsilon^2)\sim 0.1$ corrections to the operator are also 
contained within this conservative uncertainty  estimate. We display the band obtained for the $R_0=1.0$~fm
cutoff which encompasses the N$^2$LO calculations with both cutoffs
($R_0=1.0,1.2$~fm). For each panel, it is clear that a plateau in the
ratio sets in at a value $R$ depending on the scale and scheme.
Moreover, we observe from Fig.~\ref{fig:a2} that the $r=0$ value is a
conservative estimate for $a_2$ given that the statistical uncertainties
in the calculation of the two-body distributions grow as we approach
zero separation. As is evident from Fig.~\ref{fig:a2}, the GFMC values
for $a_2$ are in very good agreement with experiment~\cite{hen2012}
while the preliminary VMC results are also encouraging. 
We summarize the extracted SRC scaling factors $a_2$ of the GFMC
calculations and the comparison with experiment in Table~\ref{tab:a2}.

\begin{table}
\caption{\label{tab:a2}
Results for the SRC scaling factor $a_2$ obtained via
Eq.~(\ref{eq:gfmca2}) from GFMC calculations of $A=2,3,4$ nuclei based
on chiral N$^2$LO interactions (for cutoffs $R_0 = 1.0$ and $1.2$~fm)
and the AV18+UIX potentials. The uncertainties quoted for the N$^2$LO
interactions include the uncertainty estimated from the truncation of
the chiral expansion  added in quadrature to the GFMC statistical
uncertainties. 
}
\begin{ruledtabular}
\begin{tabular}{lccl}
& N$^2$LO ($R_0=1.0-1.2$~fm) & AV18+UIX & Exp.~\cite{hen2012} \\ 
\hline
$^3$H & $2.1(2)-2.3(3)$ & $2.0(4)$ & \\
$^3$He & $2.1(2)-2.1(3)$ & $2.0(4)$ & $2.13(4)$ \\ 
$^4$He & $3.8(7)-4.2(8)$ & $3.4(3)$ & $3.60(10)$
\end{tabular}
\end{ruledtabular}
\end{table}

{\bf Summary and outlook:} We have shown that the linear relation
between the magnitude of the EMC effect at intermediate $x$ and the SRC
scaling factor $a_2$ is a natural consequence of scale separation and
have derived this result using EFT. We have also
computed  $a_2$  for $^3$He and $^4$He  using the GFMC method with both
chiral and Argonne-Urbana potentials to confirm the scheme and scale
independence. 

GFMC calculations with chiral interactions for $^9$Be, $^{12}$C
and other light nuclei will allow further tests of the EFT understanding
of these phenomena. In the case of  $^9$Be, it would be especially
interesting to confirm whether $a_2$ is determined by local instead of
global nuclear density \cite{Seely:2009gt}. It would also be very
insightful to complete our theoretical understanding of the EMC-SRC
relation by computing the $C(x)$ coefficient in Eq.~(\ref{EMC-SRC}) from
lattice QCD calculations of $f_2(x)$
 from the
deuteron~\cite{Detmold:2004kw,Ji:2013dva,Lin:2014zya}.

The EFT approach to the partonic structure of nuclei has broader
applicability than to the isoscalar structure that we have discussed
above. For the $F_3(x,Q^2)$ structure function that is accessible in
weak-current DIS, EFT predicts a  relation analogous to Eq.~(\ref{F2A})
with $F_2$  replaced by $F_3$, and $g_2$ replaced by an
isospin-dependent nuclear matrix element. The resulting analogue of
Eq.~(\ref{EMC-SRC}) is also expected to hold. The generalization to
spin-dependent parton structure and to generalized parton distributions
\cite{[{See, e.g., }][]GPD} is similarly straight forward. EFT could
also shed light on whether a plateau of $\sigma_A/\sigma_{^3\text{He}}$
for $2<x<3$ exists, which is still inconclusive experimentally
\cite{Egiyan:2005hs,Fomin:2011ng,Higinbotham:2014xna}.

\begin{acknowledgments}
\textit{Acknowledgments:} 
We thank Or Hen, Evgeny Epelbaum, Dick Furnstahl, Martin Hoferichter, Dean Lee,
Ulf-G. Mei\ss ner, Jerry Miller, and Akaki Rusetsky for useful comments and
discussions. This work was supported in part by MOST of Taiwan under
Grant  Nos. 102-2112-M-002-013-MY3 and 105-2918-I-002-003, DOE
Early Career Research Award DE-SC0010495, DOE Grant No. DE-SC0011090,
the ERC Grant No.\ 307986 STRONGINT, the MIT MISTI program and the Kenda Foundation of Taiwan.
The computations were performed at NERSC, which is supported by the US
Department of Energy under Contract No.~DE-AC02-05CH11231, and on the
Lichtenberg high performance computer of the TU Darmstadt.
\end{acknowledgments}

\bibliography{supplement}

\section{Supplemental Material}

\section{A. Details on Effective Field Theory}

In Weinberg's power counting scheme, the typical nucleon momenta  $|{\bf q}|$
are counted as $\mathcal{O}(\epsilon)$, where $\epsilon$ is the ratio of the soft
to hard scale, while their energies $q^0$ are $\mathcal{O}(\epsilon^2)$.
Two-nucleon contact operators $(N^{\dagger }N)^{2}$ are counted as
$\mathcal{O}(\epsilon^0)$, while the three-body
contact operator $(N^{\dagger }N)^{3}$ is counted as $\mathcal{O}(\epsilon^3)$,
both according to their mass dimension.

This counting can be applied to Eq.~(6) of the Letter where the leading twist
quark operator is matched to hadronic operators \cite{Chen:2004zx}
\begin{eqnarray} \label{1}
\mathcal{O}^{\mu _{0}\ldots \mu _{n}} &\to& 
:\langle x^{n}\rangle _{N}
M^n v^{(\mu_{0}}\cdots v^{\mu _{n})}N^{\dagger }N
\left[ 1+\alpha _{n} 
N^{\dagger }N\right] \nonumber \\
&+& \langle x^{n}\rangle _{\pi} \pi^{\alpha} i\partial^{(\mu _{0}}
\cdots i\partial^{\mu _{n})} \pi^{\alpha}+\ldots: .  
\end{eqnarray}
We will focus on the tensor component with all $\mu_i=0$. Since $v^0 = 1$, 
the $v^{(\mu_{0}}\cdots v^{\mu _{n})} ( N^{\dagger }N)$ operator is 
$\mathcal{O}(\epsilon^{-3})$, $v^{(\mu_{0}}\cdots v^{\mu _{n})} ( N^{\dagger }N) ^{2}$
is $\mathcal{O}(\epsilon^0)$ while $v^{(\mu_{0}}\cdots v^{\mu _{n})} ( N^{\dagger }N) ^{3}$
is $\mathcal{O}(\epsilon^3)$.
The one derivative operator $\partial^{(\mu_{0}}v^{\mu_1}\cdots v^{\mu _{n})} ( N^{\dagger }N) $
is $\mathcal{O}(\epsilon^{-1})$, but its net effect is to replace $p^0$ in Eq.~(4) of the Letter
from $AM$ to $M_A$. 

The two derivative operator 
$\partial^{(\mu_{0}}\partial^{\mu_1} v^{\mu_2}\cdots v^{\mu _{n})} ( N^{\dagger }N)$
is $\mathcal{O}(\epsilon)$, and it  
can ``spill" $q_N(x)$ to $x>1$. This is the Fermi-motion effect. Although it is higher
order than the two-body operator, if $\tilde{q}_2(x)$ is very small when $x$ is just above one,
then the Fermi-motion effect could become larger and explain why the $a_2$ plateau
only sets in at $x \gtrsim 1.5$. 
It is important to note that in EFT off-shell effects that enter through Fermi motion can be absorbed
into two-body operators through a field redefinition \cite{Politzer:1980me,Arzt:1993gz}.
Therefore the separation between ``Fermi motion" and ``two-body effects" are meaningful
only after the theory is clearly specified.

The pion one-body operator 
$\pi^{a} i\partial^{(\mu _{0}}\cdots i\partial^{\mu _{n})} \pi^{a}$
inserted in the one-pion-exchange diagram contributes at 
$\mathcal{O}(\epsilon^{n-1})$.  Since $\langle x^{n}\rangle _{\pi}=0$
for even $n$ due to charge conjugation symmetry, the $n=1$ pion
operator enters at $\mathcal{O}(\epsilon^{0})$, but for higher $n$ the
contributions either vanish or are higher order compared with the
other operators in Eq.~(\ref{matching}) of the Letter. This pion contribution 
is proportional to $\delta(x)/x$ at $\mathcal{O}(\epsilon^{0})$. It breaks the
factorization of the $x$ and $A$ dependence in nuclear PDFs but only at $x=0$. So 
Eq.~(\ref{qA}) of the Letter still holds for $x\ne 0$.

All the other operators in the matching are power counted and found to be
higher order than $\epsilon^{0}$.   
In Ref. [43], a large pionic effect was found by computing
the pion number in a nucleus. This result is not in contradiction to our discussion
above, since the pion number operator $\pi^{\dagger} \pi$ is an
$\mathcal{O}(\epsilon^{-2})$ effect, but this operator does not appear in the
matching of Eq.~(\ref{1}).

\section{B. Mixed Estimates in GFMC Calculations}

To calculate the two-body distributions $\rho_{2,\mathbf{1}}(r)$ defined
in Eq.~(\ref{eq:2bddist}) of the Letter, we employ the GFMC
method, starting from a variational Monte Carlo (VMC) calculation for
the best possible trial wave function $\ket{\Psi_T}$ that minimizes the
expectation value $\tfrac{\ev{H}{\Psi_T}}{\braket{\Psi_T}}$.
The GFMC propagation, by acting with the imaginary-time propagator
$e^{-H\tau}$, evolves this trial state to the many-body ground state of
the Hamiltonian $|\Psi_T\rangle\to|\Psi_0\rangle$.
Ideally, one would propagate both trial wave functions in a given
expectation value to large imaginary time:
$\ev{\mathcal{O}(\tau)}=\tfrac{\ev{\mathcal{O}}{\Psi(\tau)}}
{\braket{\Psi(\tau)}}$, with $\ket{\Psi(\tau)}=e^{-H\tau}\ket{\Psi_T}$,
such that
\begin{equation}
\lim_{\tau\to\infty}\ev{\mathcal{O}(\tau)}\to
\ev{\mathcal{O}}=\frac{\ev{\mathcal{O}}{\Psi_0}}{\braket{\Psi_0}}\,.
\end{equation}
While such an expectation value is possible to compute in principle, it
is difficult in practice for spin- and isospin-dependent
operators and more so for momentum-dependent operators.

Instead, an approximation involving a so-called ``mixed estimate'' is
employed:
\begin{equation}
\label{eq:mxdest}
\langle\mathcal{O}\rangle=\frac{\langle\Psi_0|\mathcal{O}|\Psi_0\rangle}
{\langle\Psi_0|\Psi_0\rangle}\approx2
\frac{\langle\Psi_0|\mathcal{O}|\Psi_T\rangle}
{\langle\Psi_0|\Psi_T\rangle}
-\frac{\langle\Psi_T|\mathcal{O}|\Psi_T\rangle}
{\langle\Psi_T|\Psi_T\rangle}\,,
\end{equation}
where the term
$\tfrac{\mel{\Psi_0}{\mathcal{O}}{\Psi_T}}{\braket{\Psi_0}{\Psi_T}}$ is
the mixed estimate.
\Cref{eq:mxdest} is obtained by assuming
$\ket{\Psi_T}=\ket{\Psi_0}+\ket{\delta\Psi}$, with $\ket{\delta\Psi}$
small, and keeping terms only to $\mathcal{O}(\delta\Psi)$.
This approximation has been tested thoroughly for many operators.
For more details see Ref.~\cite{Pudliner:1997ck}.

\begin{figure}[t]
\includegraphics[width=\columnwidth]{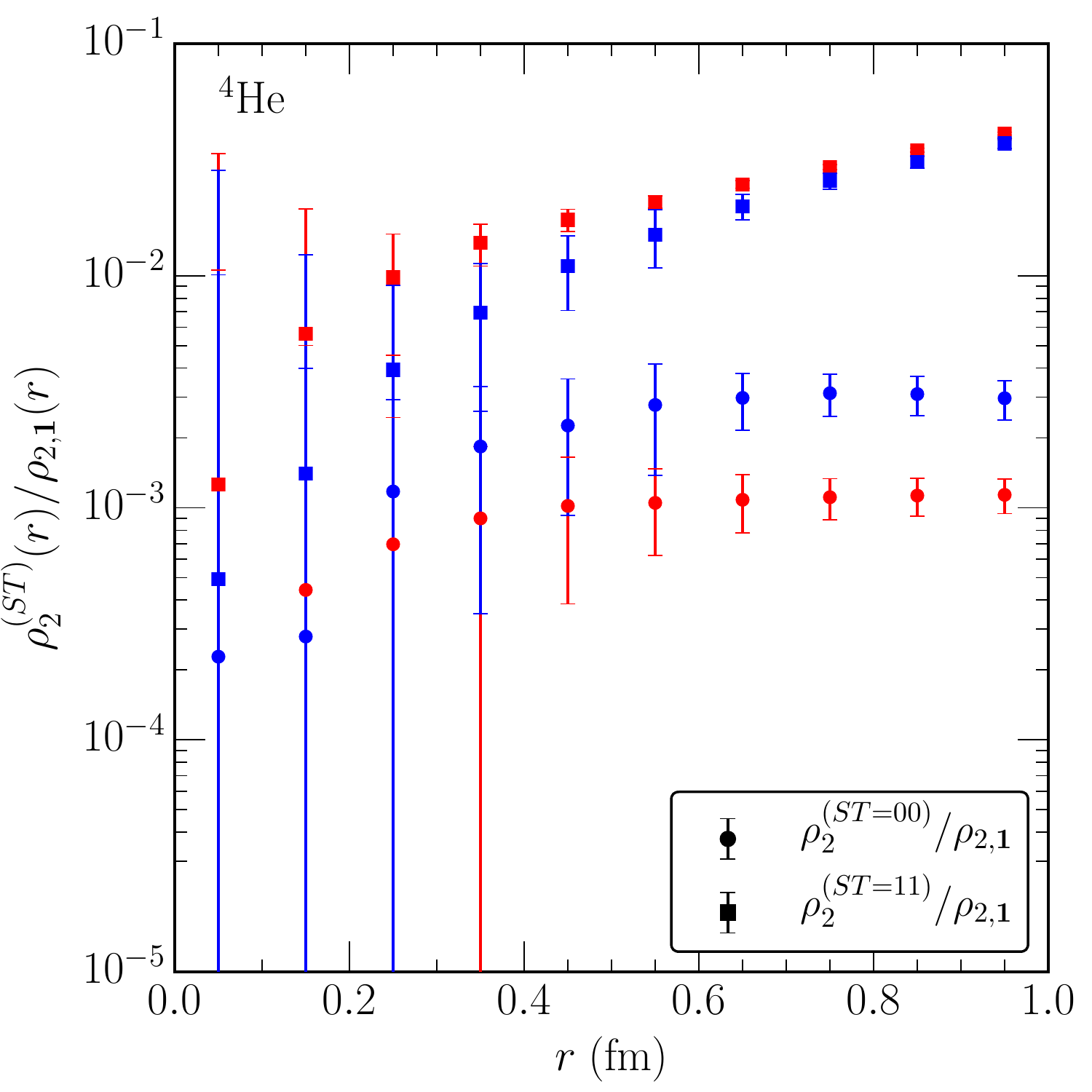}
\caption{\label{fig:he4rat}
Ratio of $(ST)=(00)$ and $(11)$ two-body distributions to the
leading SU(4)-symmetric combination $\rho_{2,\mathbf{1}}(r)$ in
\isotope[4]{He} calculated using the GFMC method.
Blue points are for the \nxlo{2} chiral interactions with $R_0=1.0$~fm.
Red points are for the AV18+UIX potentials.
The error bars here represent the GFMC statistical uncertainties.}
\end{figure}

So long as the extrapolation, (the difference between the mixed estimate
and the variational estimate), is relatively small, then one obtains a
robust value of the observable independent of the starting trial wave
function.
We have checked in all cases in this Letter that the difference between
the mixed estimate and the variational estimate is no more than 5\%
(15\%) of the mixed estimate for the chiral EFT
($\text{AV18}+\text{UIX}$) interactions.
\begin{figure}[t!!]
\includegraphics[width=\columnwidth]{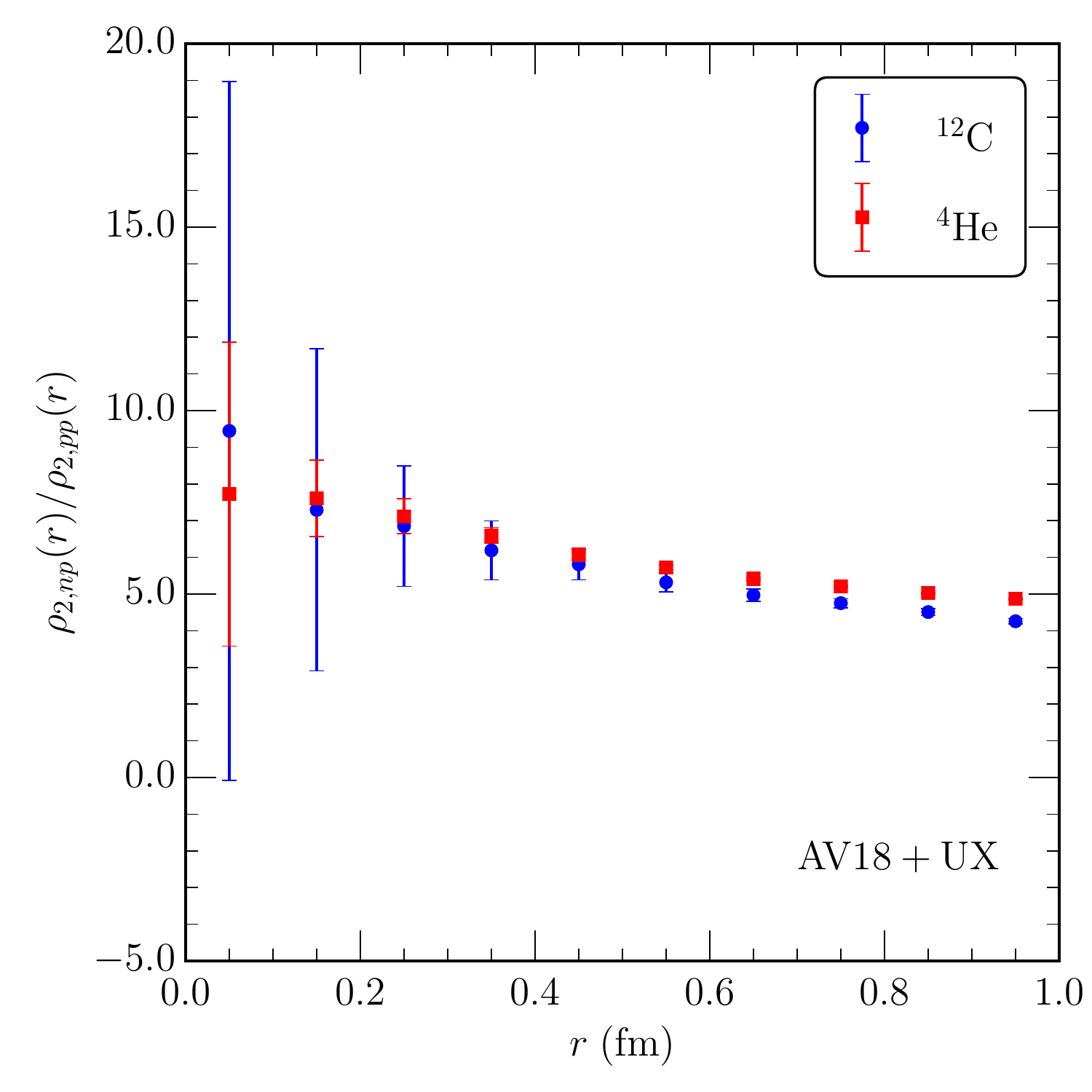}
\caption{\label{fig:he4c12nptopp}
Ratio of $np$ to $pp$ two-body distributions for \isotope[4]{He} (red
squares) and \isotope[12]{C} (blue circles) extracted from VMC
calculations with the $\text{AV18}+\text{UX}$ potentials~\cite{2bdens}.
The error bars here represent the statistical VMC uncertainties.}
\end{figure}

\section{C. Details on Two-Body Distributions}

In Fig.~\ref{fig:a2} of the Letter, we show that when $r$ is smaller than the
resolution scale $R$, then the $r$ dependence in the ratio drops out.
This implies that the physics of $a_2$ is not governed by distance
scales shorter than $R$, which cancel in the ratio of two-body
distributions.
Instead, it is governed by physics at a scale larger than $R$.
Because the physics of the $a_2$ ratio is governed by a scale larger
than $R$, higher partial-wave contributions are naturally suppressed.
In~\cref{fig:he4rat}, we show that the non~$S$-wave two-body
distributions $\rho_2^{(ST)}(r)$ for the two-body spin/isospin
$(ST)=(00)$ (circles) and $(11)$ (squares) channels in \isotope[4]{He}
are indeed much smaller than the leading SU(4)-symmetric $S$-wave
combination $\rho_{2,\mathbf{1}}(r)$, suppressed by at least an order of
magnitude.
Note that we should not expect the ratios in~\cref{fig:he4rat} to be
constant in $r$ at small $r$ since those operators renormalize
differently.

\vspace{5em}
The $ST$ distributions are defined in terms of $S$ and $T$
projectors for a pair of nucleons $\ket{ij}$:
\begin{subequations}
\begin{align}
\mathcal{P}^{(S=0)}_{ij}&\equiv\frac{1}{4}(1-\sdots{i}{j})\,,\\
\mathcal{P}^{(S=1)}_{ij}&\equiv\frac{1}{4}(3+\sdots{i}{j})\,,
\end{align}
\end{subequations}
and similarly for $\mathcal{P}^{(T)}_{ij}$, with $\sigma\to\tau$.
Then,
\begin{equation}
\rho_2^{(ST)}(r)\equiv\frac{1}{4\pi r^2}\ev{\sum_{i<j}^{A}
\mathcal{P}_{ij}^{(S)}\mathcal{P}_{ij}^{(T)}
\delta(r-r_{ij})}{\Psi_0\vphantom{\sum}}\,,
\end{equation}
with $r_{ij}\equiv|\vb{r}_i-\vb{r}_j|$.

Even though our approach is based on $N_c$ counting instead of $np$-pair
dominance, we also show the ratio of $np$ to $pp$ pairs in
\isotope[4]{He} and \isotope[12]{C} in~\cref{fig:he4c12nptopp}.
These results are extracted from VMC calculations with the
$\text{AV18}+\text{UX}$ potentials~\cite{2bdens}.
The ratio ranges from 5--10 for $r<1$~fm, consistent with Fig.~3 of
Ref.~\cite{Schiavilla:2006xx}, which shows that
$\rho_{np}(q)/\rho_{pp}(q)$ ranges from 3--7 for $q>1.5$~fm$^{-1}$, with
$q$ the relative momentum between the two nucleons.
Note that this result is with the center-of-mass momentum $Q$ of the two
nucleons integrated.
For $Q=0$, the ratio is higher (see Fig.~2 of
Ref.~\cite{Schiavilla:2006xx}).

\end{document}